\documentclass[aps,twocolumn,pre,showkeys,showpacs,eqsecnum, citeautoscript]{revtex4-1}
\usepackage{graphpap}
\usepackage[dvips]{graphicx}
\usepackage[dvips]{graphics}
\usepackage{xcolor}
\usepackage{amsmath, amssymb, amsfonts, amsthm}
\usepackage{bbold}
\usepackage[pagebackref=false]{hyperref}
\usepackage{xr}
\makeatletter
\newcommand*{\addFileDependency}[1]{
  \typeout{(#1)}
  \@addtofilelist{#1}
  \IfFileExists{#1}{}{\typeout{No file #1.}}
}
\makeatother

\newcommand*{\myexternaldocument}[1]{%
    \externaldocument{#1}%
    \addFileDependency{#1.tex}%
    \addFileDependency{#1.aux}%
}

\myexternaldocument{wExtended_figures}

\begin{document}


\title{Qubits made by advanced semiconductor manufacturing}

\author{A.M.J. Zwerver$^{1}$, T. Kr\"{a}henmann$^{1}$, T.F. Watson$^{2}$, L. Lampert$^{2}$, H.C. George$^2$, R. Pillarisetty$^2$, S.A. Bojarski$^2$, P. Amin$^2$, S.V. Amitonov$^1$, J.M. Boter$^1$, R. Caudillo$^2$, D. Corras-Serrano$^2$, J.P. Dehollain$^1$, G. Droulers$^1$, E.M. Henry$^2$, R. Kotlyar$^2$, M. Lodari$^1$, F. Luthi$^2$, D.J. Michalak$^2$, B.K. Mueller$^2$, S. Neyens$^2$, J. Roberts$^2$, N. Samkharadze$^1$, G. Zheng$^1$, O.K. Zietz$^2$, G. Scappucci$^1$, M. Veldhorst$^1$, L.M.K. Vandersypen$^{1,*}$, J.S. Clarke$^{2,*}$}

\date{\today}

\begin{abstract}
\emph{%
$^1$ QuTech and Kavli Institute of Nanoscience, Delft University of Technology, 2600 GA Delft, The Netherlands\\%
$^2$ Intel Components Research, Intel Corporation, 2501 NW 229th Avenue, Hillsboro, OR, USA \\}
$^*$ Corresponding authors: l.m.k.vandersypen@tudelft.nl; james.s.clarke@intel.com\\

Full-scale quantum computers require the integration of millions of quantum bits~\cite{campbell_roads_2017, wecker_gate-count_2014}. The promise of leveraging industrial semiconductor manufacturing to meet this requirement has fueled the pursuit of quantum computing in silicon quantum dots. However, to date, their fabrication has relied on electron-beam lithography~\cite{veldhorst_addressable_2014,maurand_cmos_2016,petit_spin_2018,yang_operation_2020,harvey-collard_spin-orbit_2019, xue_benchmarking_2019,yoneda_quantum-dot_2018,watson_programmable_2018,zajac_resonantly_2018,andrews_quantifying_2019, wu_two-axis_2014, li_flexible_2020} and, with few exceptions~\cite{maurand_cmos_2016,harvey-collard_spin-orbit_2019, li_flexible_2020}, on academic style lift-off processes. Although these fabrication techniques offer process flexibility, they suffer from low yield and poor uniformity. An important question is whether the processing
conditions developed in the manufacturing fab environment to enable high yield, throughput, and uniformity of transistors are suitable for quantum dot arrays and do not compromise the delicate qubit properties. Here, we demonstrate quantum dots hosted at a $^{28}$Si/$^{28}$SiO$_2$ interface, fabricated in a 300~mm semiconductor manufacturing facility using all-optical lithography and fully industrial processing~\cite{pillarisetty_qubit_2018}. As a result, we achieve nanoscale gate patterns with remarkable homogeneity. The quantum dots are well-behaved in the multi-electron regime, with excellent tunnel barrier control, a crucial feature for fault-tolerant two-qubit gates. Single-spin qubit operation using magnetic resonance reveals relaxation times of over $1$ s at $1$~Tesla and coherence times of over $3$~ms, matching the quality of silicon spin qubits reported to date~\cite{veldhorst_addressable_2014,maurand_cmos_2016,harvey-collard_spin-orbit_2019,petit_spin_2018,yang_operation_2020,xue_benchmarking_2019,yoneda_quantum-dot_2018,watson_programmable_2018,zajac_resonantly_2018,andrews_quantifying_2019, wu_two-axis_2014}. The feasibility of high-quality qubits made with fully-industrial techniques strongly enhances the prospects of a large-scale quantum computer.

\end{abstract}

\maketitle
\newpage

The idea of exploiting quantum mechanics to build computers with computational powers beyond the abilities of any classical device has gathered momentum since the 1980's \cite{feynman_simulating_nodate}. However, in order for full-fledged quantum computers to become a reality, they need to be fault-tolerant, i.e. errors from unavoidable decoherence must be reversed faster than they occur~\cite{campbell_roads_2017}. The most promising architectures require a scalable qubit system of individually-addressable qubits with a gate fidelity over $99\%$ and tunable nearest-neighbour couplings~\cite{bravyi_quantum_1998, fowler_surface_2012}. 

Spin qubits in gate-defined quantum dots (QDs) offer great potential for quantum computation due to their small size and relatively long coherence times \cite{loss_quantum_1998, zwanenburg_silicon_2013,vandersypen_quantum_2019}.
Single-qubit gate fidelities over 99.9\%~\cite{yoneda_quantum-dot_2018,yang_silicon_2019} as well as two-qubit gates~\cite{veldhorst_two-qubit_2015,zajac_resonantly_2018,huang_fidelity_2019, xue_benchmarking_2019, hendrickx_fast_2020}, algorithms~\cite{watson_programmable_2018}, conditional teleportation~\cite{qiao_conditional_2020}, three-qubit entanglement~\cite{takeda_quantum_nodate} and four-qubit universal control~\cite{hendrickx_four-qubit_2020} have already been demonstrated. Moreover, silicon spin qubits have been operated at relatively high temperatures of 1-4 K~\cite{petit_universal_2020, yang_operation_2020}, where the higher cooling power enables scaling strategies with integration of control electronics~\cite{vandersypen_interfacing_2017, veldhorst_silicon_2017, li_crossbar_2018, pauka_cryogenic_2019, xue_cmos-based_2020}.

A major advantage of silicon spin qubits is that they could leverage decades of technology development in the semiconductor industry. Today, industry is able to make uniform transistors with gate lengths of several tens of nanometers and spaced apart by $34$~nm (fins) to $54$~nm (gates), feature sizes that are well below the $193$~nm wavelength of the light used in the lithography process~\cite{auth2017}. This engineering feat and the high yield that allows integrated circuits containing billions of transistors to function, are enabled by adhering to strict design rules and by advanced semiconductor manufacturing techniques such as multiple patterning for pitch doubling, subtractive processing, chemically-selective plasma etches, and chemical mechanical polishing (CMP)~\cite{zant2014}. While these processing conditions are more intrusive than the metal lift-off processing conditions typically used on academic devices, they will be key to achieving the extremely high yield necessary for the fabrication of thousands or millions of qubits in a functional array.

A quantum dot device bears a strong similarity to a transistor, taken to the limit where the gate above the channel controls the flow of electrons one at a time~\cite{averin86}. In linear qubit arrays, the transistor gate is replaced by multiple gates, used to shape the potential landscape of the channel into multiple potential minima (quantum dots), to control the occupation of each dot down to the last electron, and to precisely tune the wavefunction overlap (tunnel coupling) of electrons in neighbouring dots~\cite{van_der_wiel_electron_2002}. 
In addition, qubit devices commonly rely on integrated nearby charge sensors to enable single-shot spin readout and high-fidelity initialisation~\cite{hanson_spins_2007, vandersypen_quantum_2019}. A first key question is then whether the reliable but strict design rules of industrial patterning can produce suitable qubit device layouts. A separate consideration is that qubit coherence is easily affected by microscopic charge fluctuations from interface, surface and bulk defects. Therefore, a second key question is whether the coherence properties of the qubits survive the processing conditions needed to achieve high yield and uniformity. Although the first qubits in quantum dots fabricated on wafer-scale in industrial foundries have been presented~\cite{maurand_cmos_2016, chanrion_charge_2020}, they rely on electron-beam lithography and avoid CMP for the active device area. CMP requires a uniform metal density across the wafer, which introduces its own complexities for quantum dot devices, due to the large amount of floating metal and added capacitance.

In this letter, we demonstrate optically-patterned quantum dots and qubits, made in a state-of-the-art 300~mm wafer process line, similar to those used for commercial advanced integrated circuits.

A dedicated mask set based on 193~nm immersion lithography is created for patterning quantum dot arrays of various lengths, as well as a number of test structures, such as transistors of various sizes and Hall bars. These test structures allow us to directly extract important metrics at both room temperature and low temperature such as mobility, threshold voltage, subthreshold slope and interface trap density. Analysed together, these metrics give us understanding of the gate oxide and contact quality along with the electrostatics to help troubleshoot process targeting~\cite{pillarisetty_qubit_2018}.

\begin{figure}[h!]
    \centering
    \includegraphics[width=0.9\textwidth]{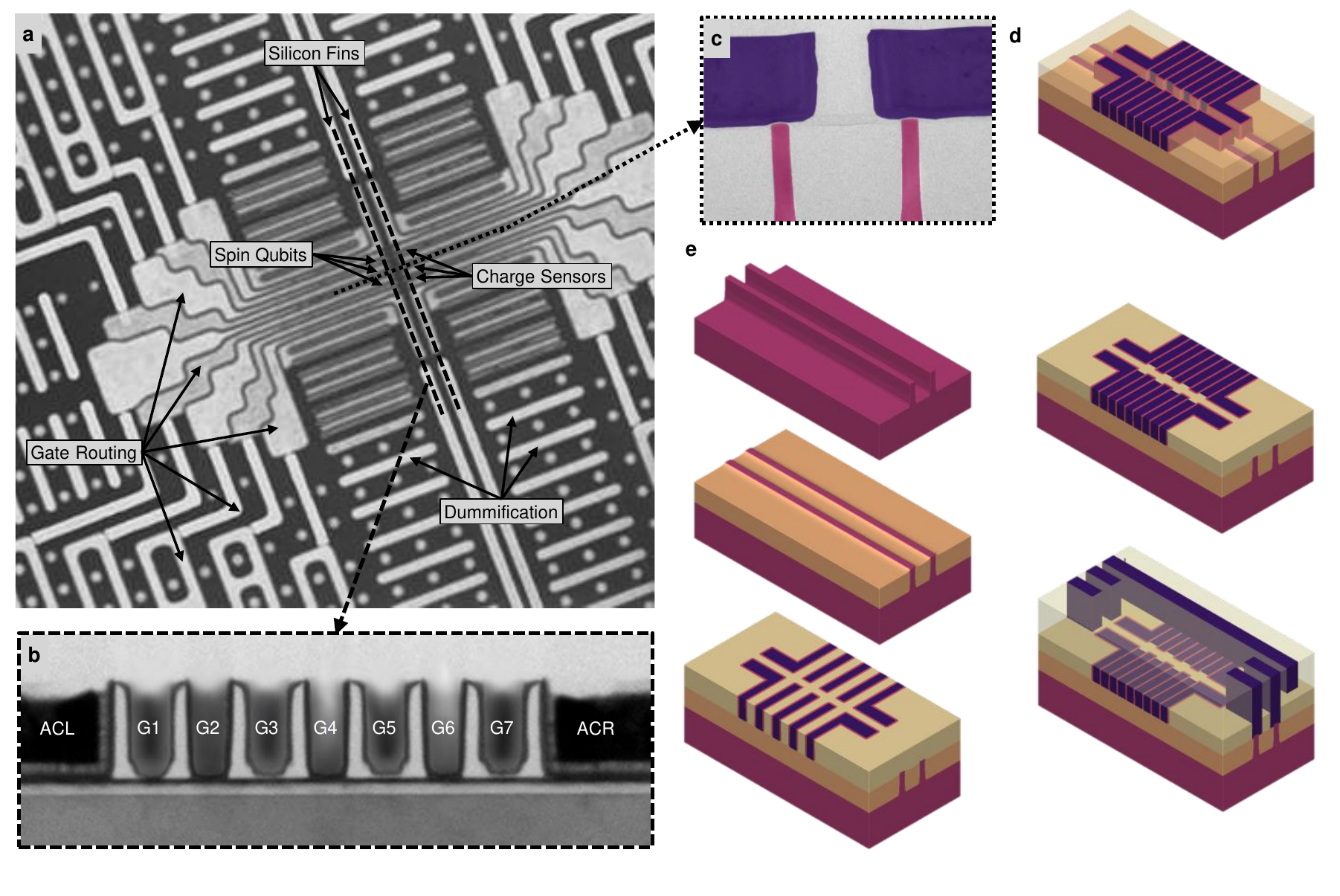}
    \caption{\textbf{Industrially fabricated quantum dot devices.} \textbf{a,} HAADF-STEM image of a typical device. The active region consists of two parallel fins; one hosts the qubits and the other hosts the sensing dot. The fan-out of the gates is clearly visible, as are many additional metallic structures (called dummification) needed to maintain a roughly constant density of metal on the surface, which ensures homogeneous polishing on a wafer scale. \textbf{b,} TEM image along a Si fin, showing 7 finger gates to define the quantum dot array and two accumulation gates to induce reservoirs connecting to the n-type implants that serve as Ohmic contacts (outside the image). \textbf{c,} False-coloured TEM image perpendicular to the Si fins, showing the fins and the gates on top. \textbf{d,} Schematic of the active region of the device. \textbf{e,} Schematic of the process steps used to fabricate the devices as explained in the main text.}
    \label{fig:figure1}
\end{figure}

As in current complementary metal-oxide-semiconductor (CMOS) transistors, the active region of these quantum dot devices consists of a fin etched out of the silicon substrate~\cite{auth2017}. Nested top-gates with a pitch of $50$~nm, separated from the fin by a composite SiO$_2$/high-k dielectric, are used to form and manipulate quantum dots. Figure~\ref{fig:figure1}{a} shows a high-angle annular dark-field scanning transmission electron microscopy (HAADF-STEM) image of the active device area. A cross-section transmission electron microscopy (TEM) image along a fin with quantum dot gates is shown in Fig.~\ref{fig:figure1}{b}.  N-type ion implants on both ends of the fin, well separated from the active region, serve as Ohmic contacts to the fins. We pattern two such linear quantum dot arrays, separated by 100~nm (see Fig.~\ref{fig:figure1}{c} for a TEM image across both fins). In our experiments, we use a quantum dot in one array as a charge sensor for the quantum dots and qubits in the other array. A 3D device schematic is shown in Fig.~\ref{fig:figure1}{d}. 

The process flow starts from a conventional transistor flow but is adapted to fabricate two sets of gates in successive steps using a combination of 300~mm optical lithography, thin film deposition, plasma etch, and chemical mechanical polish processes. The main steps are illustrated in Fig.~\ref{fig:figure1}{e}.  First, the fins are defined in a Si substrate.  The area between the fins is filled in with a shallow trench isolation dielectric material and polished.  The first gate layer (with even numbers) is defined using an industry standard replacement metal gate process~\cite{natarajan_32nm_2008, auth_22nm_2012, mistry_logic_2007} and the second gate layer (with odd numbers) is then formed adjacent to the first gate layer. Finally, a contact layer is patterned to enable routing to bond pads, as well as ohmic and gate contacts.  In the devices intended for coherent experiments (discussed in Fig.~\ref{fig:figure4}), a metal wire that shunts a coplanar stripline (connected to an on-chip coplanar waveguide) is placed parallel to the fins in the contact layer (Extended Data Fig.~2). The samples used only for quantum dot formation are fabricated on natural silicon substrates, whereas the samples used for qubit readout and manipulation are fabricated on an isotopically enriched $^\mathrm{28}$Si epilayer with a residual $^\mathrm{29}$Si concentration of 800 ppm~\cite{sabbagh_quantum_2019}. This reduces the hyperfine interaction of the qubits with nuclear spins in the host material and thus increases the qubit coherence~\cite{zwanenburg_silicon_2013}.

A single 300~mm wafer contains more than $10,000$ quantum dot arrays of various lengths, with up to 55 gates per fin. The impact of the industrial fabrication is immediately clear when we compare the TEM image of a 7-gate device in Fig.~\ref{fig:figure1}{b} to a TEM image of an academic Si-MOS device (see Extended Data Fig.~1); the gate uniformity is remarkable and there are no traces of residual metallic pieces. This uniformity is reflected in e.g. the threshold voltage of individual gates within quantum dot arrays, as well as across quantum dot arrays throughout the wafer, measured at room temperature (see Extended Data Fig.~3).

All quantum dot and qubit measurements are carried out in a dilution refrigerator, operated at base temperature. The measurements have been performed on a plethora of different devices from different process flow generations, measured in three different dilution refrigerators in different laboratories. Samples of the same generation taken from anywhere across the wafer show very reproducible quantum dot behaviour.
 
\begin{figure}[h!]
    \centering
    \includegraphics[width=0.9\textwidth]{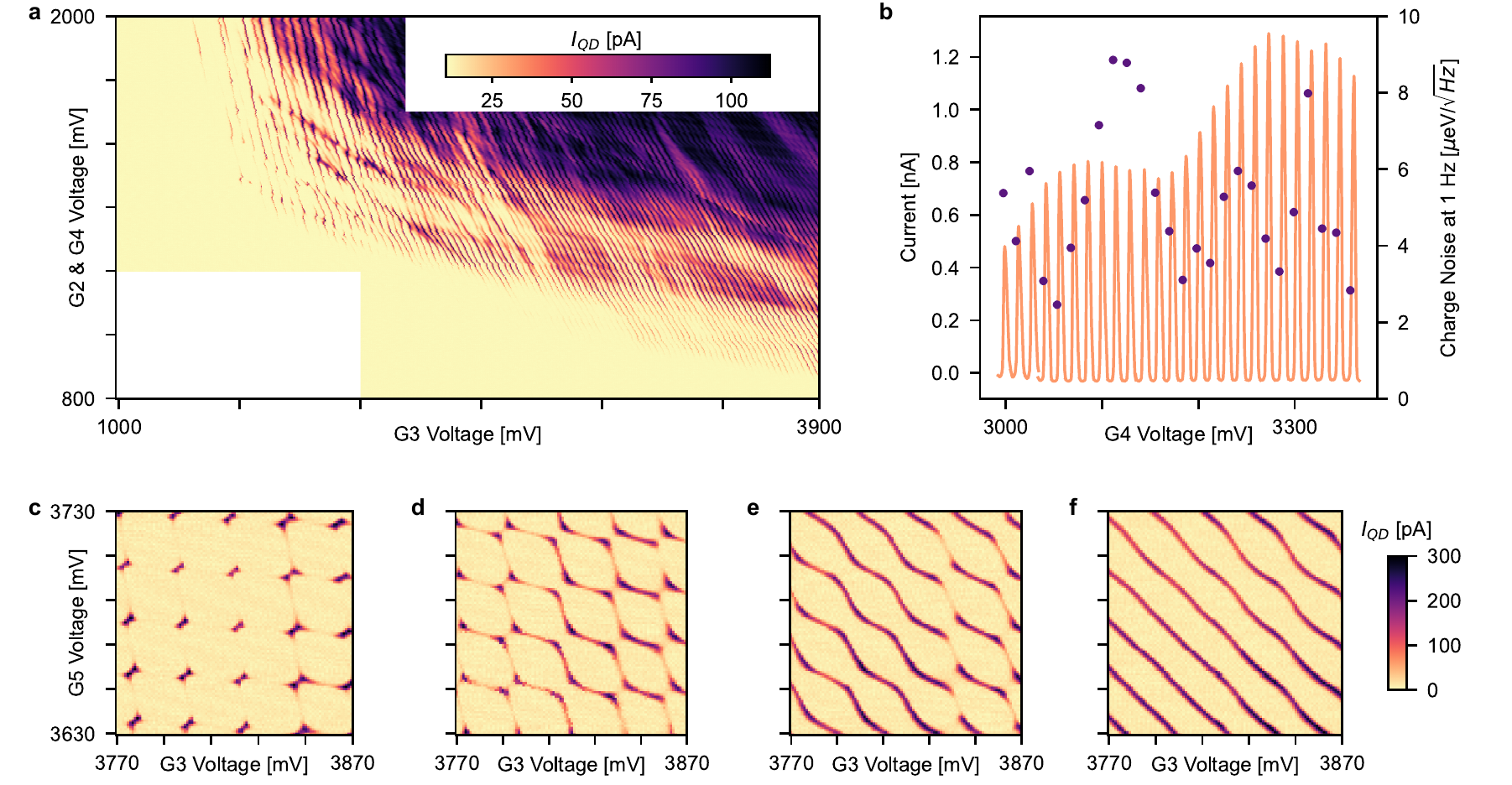}
    \caption{\textbf{Tunable single and double quantum dots} \textbf{a,} Charge stability diagram for a single QD measured via electron transport. \textbf{b,} Coulomb blockade peaks in the multi-electron regime (orange line) and the power spectral density at 1 Hz of the quantum dot potential fluctuations measured at the flank of each peak (purple dots). \textbf{c-f,} Charge stability diagrams of a double quantum dot formed under gates G3 and G5. The gate voltage on G4 is gradually increased (G4 is $1245$, $1308$, $1353$ and $1398$~mV from \textbf{c} to \textbf{f}), showing good control over the interdot tunnel coupling.}
    \label{fig:figure2}
\end{figure}

We form quantum dots by tuning the gates individually to define a suitable potential landscape (dots can be controllably formed below any of the inner 5 gates). Figure~\ref{fig:figure2}{a} shows a typical result, where we measure the current through a QD while sweeping three gate voltages, one of which mostly controls the electrochemical potential (labeled G3 in Fig.~\ref{fig:figure1}{b}) of the quantum dot and the others (G2 and G4) mostly control the tunnel barriers. Starting at a G3 gate voltage of 3.9~V, we count more than 80 lines, known as Coulomb peaks, separating regions with a stable number of electrons on the QD. This number is changed by exactly one when passing a Coulomb peak. While the Coulomb peaks are parallel and evenly spaced at a high G3 voltage ($>$ 3~V), they become irregular and also further separated towards the few-electron regime (around 1.5~V on G3). Such irregular behaviour is characteristic of Si-MOS devices, due to their close proximity to the dielectric interface~\cite{veldhorst_addressable_2014, maurand_cmos_2016, petit_spin_2018, yang_operation_2020, veldhorst_two-qubit_2015, huang_fidelity_2019,harvey-collard_spin-orbit_2019, li_flexible_2020}.

Analysing these first-generation dots through so-called Coulomb diamonds~\cite{hanson_spins_2007}, gives an average charging energy of $8.4\pm0.2$~meV (all error bars are 1$\sigma$ from the mean) per dot in the multi-electron regime (see Extended Data Fig.~4). This indicates the dot diameter to be on the order of $45$~nm, which is about the width of a single gate. Next, we measure charge noise in the multi-electron regime, by measuring the current fluctuations at a fixed gate voltage, on the flank of a Coulomb peak. The power spectral density shows a 1/$f$ slope that is characteristic of charge noise in solid-state devices~\cite{paladino_1_2014}. The charge noise amplitude is in the range of $1 - 10$~$\mu$eV/$\sqrt{\mathrm{Hz}}$ at $1$~Hz, with some variation between Coulomb peaks (Fig.~\ref{fig:figure2}{b}). These are common charge noise values in Si-MOS QD samples~\cite{petit_spin_2018}.
 
Figure~\ref{fig:figure2}{c} shows the transport through a double QD as a function of the gate voltages that (mostly) control the electrochemical potential of each dot, G3 and G5. Characteristic points of conductance are measured, so-called triple points. At these points, the electrochemical potentials of the reservoirs are aligned with the electrochemical potentials of the left and the right dot, such that electrons can tunnel sequentially through the two dots~\cite{van_der_wiel_electron_2002}. Increasing the voltage applied to the intermediate gate G4 is expected to lower the tunnel barrier between the dots, eventually reaching the point where one large dot is formed. This behaviour is seen in Figs.~\ref{fig:figure2}{c}-{f}, as the gradual transition from triple points to single, parallel and evenly spaced Coulomb peaks. This shows the tunability of the interdot tunnel coupling in this double dot, which is advantageous for two-qubit control in such a system~\cite{zajac_resonantly_2018,yang_operation_2020,takeda_quantum_nodate,xue_cmos-based_2020}.

\begin{figure}[h!]
    \centering
    \includegraphics[width=\linewidth]{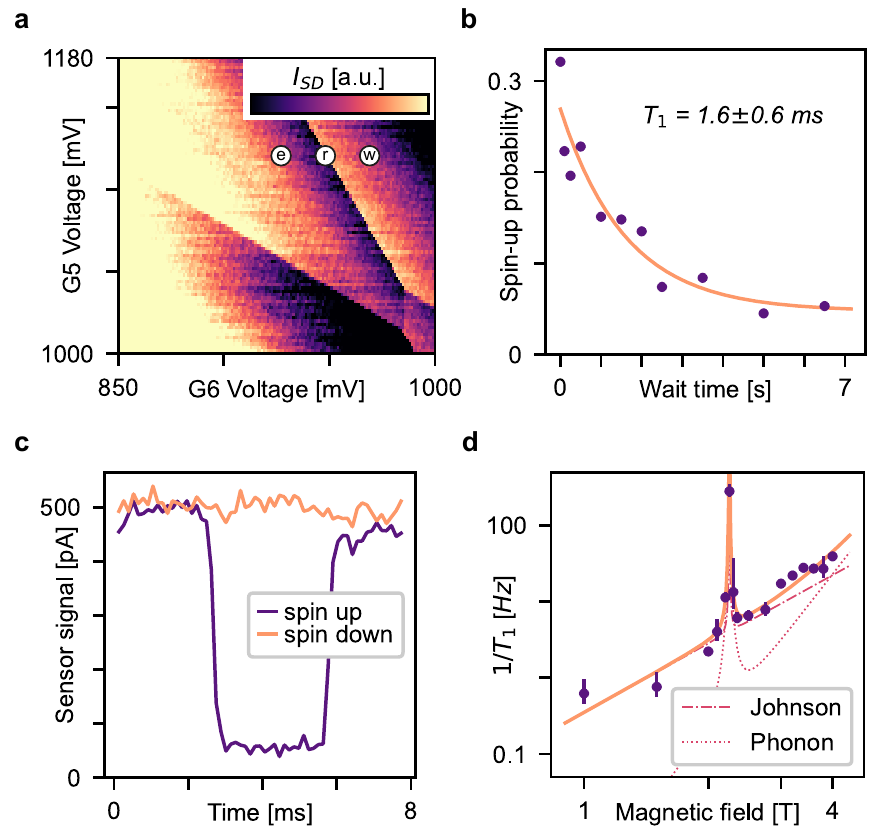}
    \caption{\textbf{Charge sensing and single-shot spin readout.} \textbf{a,} Charge stability diagram of the last-electron regime of a QD, measured with a sensing dot in the other fin. The points w, r and e refer to the wait, readout and empty stages of the gate voltage pulse. \textbf{b,} Spin-up probability as a function of load time at a magnetic field of $1$~T. The exponential fit yields a $T_1$ of $1.6\pm0.6$~s.  \textbf{c,} Real-time current through the sensing dot indicating a spin-up (purple line) and spin-down (orange line) electron, recorded with a measurement bandwidth of $3$~kHz set by an external low-pass filter. \textbf{d,} Relaxation rate ($1/T_1$) as a function of the applied magnetic field (purple dots). The relaxation rate is fitted by a model (orange line) that includes the effect of Johnson noise and phonons coupling to the spin via spin-orbit interaction. From this fit, we extract a valley splitting of $E_v=260\pm2$~$\mu$eV.}
    \label{fig:figure3}
\end{figure}

\begin{figure}[h!]
    \centering
    \includegraphics[width=\linewidth]{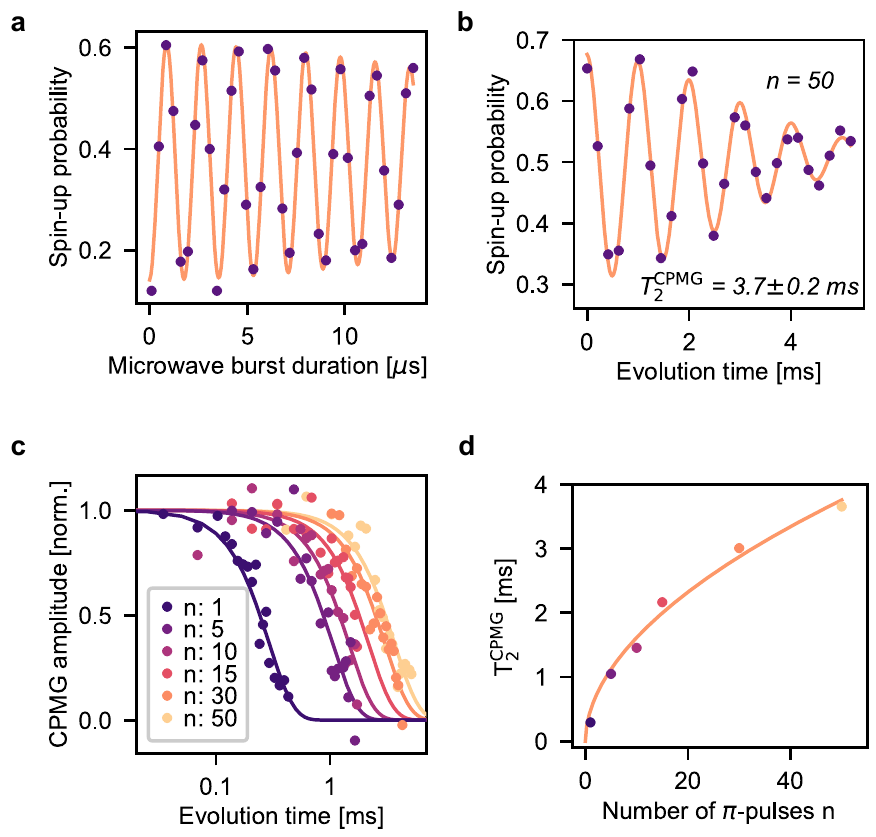}
    \caption{\textbf{An industrial silicon spin qubit} \textbf{a,} Rabi oscillations of the measured spin-up probability as a function of microwave burst duration. \textbf{b,} CPMG experiment, the measured spin-up probability as a function of the free evolution time separating 50 $\pi$ pulses, with artificial detuning. The data is fitted with  $A(\cos(\omega t+\phi)+B)\exp(-(t/T_2^{\mathrm{CPMG}})^2)+C$. The fitted CPMG coherence time $T_2^{\mathrm{CPMG}}$ is 3.7$\pm 0.2$~ms. \textbf{c,} Demodulated and normalised CPMG amplitude as a function of the total evolution time for different numbers $n$ of $\pi$-pulses. \textbf{d,} Measured coherence time $T_{2}^{\mathrm{CPMG}}$ for different numbers of CPMG $\pi$ pulses. The orange line represents a fit through the data (excluding $n=1$) following $T_2^\mathrm{CPMG}\propto n^{(\gamma/(\gamma+1))}$. We extract $\gamma=1.1\pm0.2$.}
    \label{fig:figure4}
\end{figure}

In a next step, we use a QD in one fin as a charge sensor for the charge occupation of the QDs in the other fin. This allows us to unambiguously map the charge states of the qubit dots down to the last electron~\cite{hanson_spins_2007}. A characteristic charge stability diagram showing the last electron transition is shown in Fig.~\ref{fig:figure3}{a}. The current through the sensor is measured as a function of the voltage on two gates controlling the qubit dot. In the few-electron regime, we can usually distinguish lines with several different slopes, indicating the formation of additional, spurious dots next to the intended dot. However, we consistently are able to find a clean region in the charge stability diagram with an isolated addition line corresponding to the last electron. Several iterations of geometry and material changes improved the charge sensing by orders of magnitude, resulting in a sensing step of about $500$~pA for a source-drain voltage of 500~$\mu$V. This allows single-shot readout of the spin of a single electron by means of spin-dependent tunneling and real-time charge detection (Fig.~\ref{fig:figure3}{c})~\cite{elzerman_single-shot_2004}. 

In order to define a qubit via the electron spin states, we apply a magnetic field in the $[1 0 0]$-direction, parallel to the fins, separating the spin-up and spin-down levels in energy. We apply a three-stage pulse to gate G6 to measure the spin relaxation time, $T_1$~\cite{elzerman_single-shot_2004}. We find $T_1$ exceeding $1$~s at a magnetic field of 1 Tesla (Fig.~\ref{fig:figure3}{b}). This long $T_1$ is comparable to those reported previously for silicon quantum dots~\cite{petit_spin_2018, yang_operation_2020} and indicates that the more complicated processing conditions of the 300~mm-scale fabrication do not degrade the spin relaxation time. Upon measuring $T_1$ as a function of magnetic field, we find a striking, non-monotonic dependence, which is well described in the literature and the result of the valley structure in the conduction band of silicon.  Following ~\cite{yang_spin-valley_2013, petit_spin_2018}, we fit the magnetic field dependence of the spin relaxation rate ($1/T_1$) with a model including the effect of Johnson noise and phonons inducing spin transitions mediated by spin-orbit coupling, and taking into account the lowest four valley states (Fig.~\ref{fig:figure3}{d}). The peak in the relaxation rate around $2.25$~T corresponds to the situation where the Zeeman energy equals the valley splitting energy, from which we extract a valley splitting of $260\pm2$~$\mu$eV, well above the thermal energy and qubit splitting in this system.

To coherently control the spin states, we apply an ac current to the stripline in order to generate an oscillating magnetic field at the QD~\cite{koppens_driven_2006}. Electron spin resonance occurs when the driving frequency matches the spin Larmor frequency of  $f=17.1$~GHz, which is set by the static magnetic field at the dot. By pulsing the spin-down level below the Fermi reservoir, we load the QD with a spin-down electron. We then pulse deep in the Coulomb blockade regime to manipulate the spin with microwave bursts. Finally, we pulse to the readout point and perform spin-dependent tunneling readout. The spin-up probability as a function of microwave burst duration shows clear Rabi oscillations, (Fig.~\ref{fig:figure4}{a}). We have studied Rabi oscillations in three devices; the main figures show data for qubit 1 (Q1) and the Extended figures (11, 12 and 13) show data for qubit 2 (Q2), formed on the same device, and qubit 3. As expected, the Rabi frequency is linear in the driving amplitude, reaching up to about $900$~kHz for Q2.

The spin dephasing time $T_2^*$ is measured through a Ramsey interference measurement (see Extended Data Fig.~8). Fitting this Ramsey pattern with a Gaussian-damped oscillation, yields a decay time of $T_2^*=24 \pm 6~\mu s$ when averaging data over 100~s (the error bar here refers to the statistical variation between 41 post-selected repetitions of 100~s segments). As we repeat such Ramsey measurements, we observe slow jumps in the qubit frequency. Averaging the free induction decay over 2 hours and 40 minutes still gives a $T_2^*$ of $11 \pm 2~\mu s$, see Methods section for more details.

To analyse the single-qubit gate fidelity, we employ randomised benchmarking~\cite{knill_randomized_2008} (Extended Fig.~10). A number, $m$, of random Clifford gates is applied to the qubit, followed by a gate that ideally returns the spin to either the spin-up or spin-down state. In reality, the probability to reach the target state decays with $m$ due to imperfections. The standard analysis gives a single-qubit gate fidelity of $99.0\%$ for Q1 and $99.1\%$ for Q2.
With the Rabi decay being dominated by low-frequency noise, the present combination of $T_2^*$ and Rabi frequency should allow an even higher fidelity ~\cite{veldhorst_two-qubit_2015,petit_universal_2020,yang_operation_2020}. We suspect the single-qubit gate fidelity to be limited by improper calibration. Nonetheless, the fidelity is already around the fault-tolerant threshold for the surface code~\cite{fowler_surface_2012}. 

Finally, we study the limits of spin coherence by performing dynamical decoupling by means of Carr-Purcell-Meiboom-Gill (CPMG) sequences (see Fig.~\ref{fig:figure4}{b} for the coherence decay using 50 pulses). These sequences eliminate the effect from quasi-static noise sources. Figure~\ref{fig:figure4}{c} shows the normalised amplitude of the CPMG decay as a function of evolution time for different numbers of $\pi$-pulses, $n$. By fitting these curves we extract $T_{2}^{\mathrm{CPMG}}(n)$.  We use a Gaussian decay envelope which yields distinctly better agreement than an exponential decay. The $T_{2}^{\mathrm{CPMG}}$ times are plotted as a function of $n$ in Fig.~\ref{fig:figure4}{d}. We obtain a $T_{2}^{\mathrm{CPMG}}$ of over $3.5$ ms for $n=50$ CPMG pulses, more than 100 times larger than $T_2^*$, with room for further increases through additional decoupling pulses. The CPMG data for Q1 is consistent with charge noise as the limiting mechanism (see Methods). For Q2, an additional noise mechanism is likely present. Again, all the decay timescales are comparable to the results reported earlier for $^{28}$Si-MOS devices~\cite{veldhorst_addressable_2014,veldhorst_two-qubit_2015,petit_spin_2018,yang_operation_2020,harvey-collard_spin-orbit_2019,maurand_cmos_2016}.

In summary, despite the industrial processing conditions used to fabricate the qubit samples, key performance indicators such as charge noise, the charge sensing signal, $T_1$, $T_2^*$ and $T_2^{\mathrm{CPMG}}$, are already state-of-the-art. The formation of easily tunable double dots bodes well for the implementation of two-qubit gates in this system.
Several further improvements are possible. First, the ESR stripline can be redesigned to lower resistance and dissipation by increasing the trace width up to the short and using lower resistivity materials. Furthermore, bringing the quantum dots on the inside of the stripline~\cite{koppens_driven_2006} will increase the ratio of magnetic field and (unwanted) electric fields and heating. Finally, spurious dots in the few electron regime and two-level systems can be removed by reducing the presence of material charge defects~\cite{nicollian82,schulz83}. While growth conditions for high-quality Si/dielectric interfaces have been identified, performance-limiting defects can be formed through downstream processing. Further work is ongoing to optimise the process flow and recipes (temperature budget, plasma conditions, chemical exposure, and annealing conditions) to reduce defects at the end of line.

These fabrication methods can be adapted to allow for 2D quantum dot arrays as well. Moreover, these processing steps are by default integratable with any other CMOS technology, which opens up the potential to integrate classical circuits next to the qubit chip. Eventually, industrial processing has the potential to achieve the very high quantum dot uniformity that would enable cross-bar addressing schemes~\cite{li_crossbar_2018}. The compatibility of silicon spin qubits with fully-industrial processing demonstrated here, highlights their potential for scaling and for realising a fault-tolerant full-stack quantum computer.

\clearpage
\clearpage

\clearpage

\clearpage

\clearpage

\bibliography{literature}
\bibliographystyle{naturemag}

\clearpage

\section*{Methods}

\noindent \textbf{Setup and instrumentation}\\
\noindent The measurements were performed on two different setups on different continents, setup 1 (S1, Delft) and setup 2 (S2, Hillsboro). The samples were cooled down in a dilution refrigerator, operated at the base temperature of around $10$~mK (S1: Oxford Triton dry dilution refrigerator, S2: Bluefors XLD dry dilution refrigerator). DC voltages were applied via Delft in-house built, battery-powered voltage sources (S1 and S2). The printed circuit board onto which the sample was mounted contained bias tees with a cut-off frequency of 3 Hz to allow for the application of gate voltage pulses (S1 and S2). The pulses were generated by an arbitrary waveform generator (AWG, S1: Tektronix AWG5014, S2: Zurich Instruments HDAWG). The baseband current through the sensing dot was converted to a voltage by means of a home-built amplifier, filtered through a room-temperature low-pass filter (S1: 3~kHz, S2: 1.5~kHz) and sampled by a digitiser (S1: M4i spectrum, S2: Zurich Instruments MFLI). Microwave bursts for driving ESR were generated by a vector source with an internal IQ mixer (S1 and S2:  Keysight PSG8267D), with the I and Q channels controlled by two output channels of the AWG.\\

\noindent \textbf{Charge noise measurements}\\
\noindent Each charge noise data point in Fig.~\ref{fig:figure2}{c} is obtained by recording a 140 second time trace (at 28 Hz sampling rate) of the current through the QD with the plunger gate voltage fixed at the steepest point of the Coulomb peak flank. To convert the current signal to energy, we proceed as follows. First, we convert the current to gate voltage by multiplying the data by the slope of the Coulomb peak at the operating point. Then, we multiply with the lever arm to convert from plunger gate voltage to energy. To obtain the power spectral density (PSD), we divide the data in 10 equally long segments, take the single-sided fast Fourier transform (FFT) of the segments and average these. Fitting the PSD to $A/f^\alpha$ we extract the energy fluctuations at 1 Hz ($\sqrt{A}$) for each Coulomb peak. We extract a mean value of $\alpha = 1.1\pm0.3$.\\

\noindent \textbf{Spin readout}\\
\noindent In order to read out the spin eigenstate we use energy-selective tunneling to the electron reservoir~\cite{elzerman_single-shot_2004}. The spin levels are aligned with respect to the Fermi reservoir, such that a spin-up electron can tunnel out of the QD, while for a spin-down electron it is energetically forbidden to leave the QD. Thus, depending on the spin state, the charge occupation in the QD will change. To monitor the charge state, we apply a fixed voltage bias across the sensing dot and measure the baseband current signal through the sensing dot, filtered with a low-pass filter and sampled via the digitiser. In post-analysis we threshold the sensing dot signal and  accordingly assign a spin-up or spin-down to every single shot experiment. After readout, we empty the QD to repeat the sequence. As is commonly seen in spin-dependent tunneling, the readout errors are not symmetric, which is reflected in the range of the oscillations in Figs.~\ref{fig:figure4}{a,b}.\\

\noindent \textbf{Qubit operations}\\
\noindent When addressing the qubit, we phenomenologically observe that the qubit resonance frequency shifts depending on the burst duration. The precise origin of this resonance shift is so-far unclear, but appears to be caused by heating. Similar observations have been made in recent spin qubit experiments~\cite{watson_programmable_2018, yoneda_quantum-dot_2018, xue_cmos-based_2020} that used electric-dipole spin resonance via micromagnets as the driving mechanism. To ensure a reproducible qubit frequency in the experiments, we apply an off-resonant microwave burst prior to the intended manipulation phase to saturate this frequency shift. We further investigate this frequency shift in Extended Data Figs.~6 and 7.\\

\noindent \textbf{Ramsey oscillation}\\
\noindent We observe that the qubit resonance frequency in the devices exhibits jumps of several 100~s of kHz on a timescale of 5-10 minutes. To extract meaningful results, we monitor this frequency shift throughout the experiments and accordingly discard certain data traces, such that we only take into account data acquired with the qubit in a narrow frequency window.
To illustrate the frequency shift, we show the FFT of 100 repetitions of a Ramsey interference measurement of qubit 1 (measurement time $\sim$ 2 hours 40 min) in Extended Data Fig.~8{a}, which tracks the qubit frequency over time. In order to estimate the $T_2^*$ of qubit 1, we fit each of the 100 repetitions of the Ramsey measurement (measurement time per repetition $\sim$ 100~s) and extract a $T_2^*$ value. Evidently, some of the data quality is rather poor due to the previously described frequency jumps in which case the extracted $T_2^*$ value is meaningless. We calculate the mean square error of each fit and disregard all the measurements with a high error. The average $T_2^*$ of the 41 remaining traces is $24\pm 6~\mu$s (Extended Data Fig.~8{b}). Averaging the data traces of all 41 traces and then fitting a decay curve yields a dephasing time of $16\pm2~\mu$s (Extended Data Fig.~8{c}); averaging the data of all 100 traces still gives a dephasing time of $11\pm2~\mu$s (Extended Data Fig.~8{d}).\\

\noindent \textbf{CPMG coherence measurements and power spectral density}\\
\noindent To ensure robust fitting, the CPMG sequences are applied with artificial detuning. We fit the resulting curves with a Gaussian damped cosine function: $A(\cos(\omega t+\phi)+B)\exp[-(t/T_2^{\mathrm{CPMG}})^2]+C$. If, instead of using a Gaussian decay, we leave the exponent of the decay open as a fitting parameter, we obtain values for the exponent between 2.3 and 2.6, but the use of the additional parameter results in less robust fits. The offset $B$ is included to compensate for the loss of readout visibility for long microwave burst duration. We attribute this to heating generated while driving the spin rotations. The measurement is divided into segments, each consisting of 200 single shots. Each segment includes a simple calibration part, based on which we post-select repetitions for which the spin-up probability after applying a $\pi$-pulse is above 25 percent. In this way, we can exclude repetitions where the qubit resonance frequency has shifted drastically. The remaining repetitions are averaged to obtain the characteristic decay curves for each choice of $n$, one of which is shown in Fig.~\ref{fig:figure4}{b}.
From fitting the decay curves, we extract the $T_2^\mathrm{CPMG}$ times as a function of $n$, shown in Fig.~\ref{fig:figure4}{d}. To extract the CPMG amplitude as a function of evolution time from the data, we demodulate the measured values with the parameters extracted from the fit, according to $A_\mathrm{CPMG}= (x-C)/(A(\cos(\omega t+\phi)+B))$, with $x$ the measured data. Due to experimental noise, points where the denominator is small, do not yield meaningful results. Hence, we exclude data points for which the absolute value of the expected denominator is smaller than 0.4. The extracted CPMG amplitudes are plotted in Fig.~\ref{fig:figure4}{c}. 
In a commonly used simplified framework~\cite{bylander_noise_2011,cywinski_how_nodate}, we can relate the data of Fig.~\ref{fig:figure4}{d} to a noise power spectral density of the form $S(\omega) \propto 1/\omega^\gamma$. Specifically, fitting the data to $T_2^{CPMG}(n)\propto n ^{\gamma/(\gamma+1)}$ gives $\gamma=1.1\pm0.2$. Alternatively, we can estimate $\gamma$ by fitting the noise power spectral density extracted from the individual data points in the CPMG decays~\cite{cywinski_how_nodate} in Extended Data Fig.~9(a). This analysis gives $\gamma = 1.2\pm 0.1$. Either way, the extracted power spectral density is close to the $1/f$ dependence that is characteristic of charge noise. Charge noise can affect spin coherence since the spin resonance frequency is sensitive to the gate voltage, as also reported before for Si-MOS based spin qubits~\cite{veldhorst_addressable_2014}. We next estimate how large charge noise would need to be in order to dominate spin decoherence. To do so, we extrapolate the extracted spectral density in the range between $10^3$ and $10^4$ Hz to an amplitude at $1$~Hz, which after conversion to units of charge noise gives $29\pm27$~$\mu$eV/$\sqrt{\mathrm{Hz}}$. With the caveat that this extrapolation is not very precise, we note that this value is only slightly larger than the charge noise amplitude in the multi-electron regime of $2-10$~$\mu$eV/$\sqrt{\mathrm{Hz}}$. Considering that charge noise values are typically higher in the few-electron regime, this suggests that coherence of Q1 may be limited by charge noise~\cite{cywinski_how_nodate}. For Q2, which is another qubit in the same sample, the same procedure gives an extrapolated noise at 1 Hz that is an order of magnitude larger. Possibly a two-level fluctuator is active in the vicinity of this qubit in the regime where the qubit data was taken.
\\

\section*{Acknowledgements}
We thank Luca Petit and Sander de Snoo for software support and Raymond Schouten, Raymond Vermeulen, Marijn Tiggelman, Jason Mensingh, Olaf Benningshof and Matt Sarsby for technical support. Moreover, we thank all people from the QuTech spin qubit group and from the Intel Components research group for discussions. We acknowledge financial support from Intel Corporation and the QuantERA ERA-NET Cofund in Quantum Technologies implemented within the European Union's Horizon 2020 Program.

\section*{Author contributions}
A.M.J.Z., T.K., T.F.W., L.L. and F.L. performed the quantum dot and qubit measurements. J.B., D.C.S., J.P.D., G.D., R.K., D.J.M., R.P., N.S., G.S., M.V., L.M.K.V. and J.C. designed the devices. S.A.B., H.C.G., E.M.H. and B.K.M. fabricated the devices. P.A., J.M.B., R.C., T.K., L.L., F.L., D.M., S.N., R.P., T.F.W., O.K.Z., G.Z. and A.M.J.Z. characterised the test structures and devices. M.L. characterised the Si-MOS stacks. S.V.A. contributed to the preparation of the experiments. A.M.J.Z., T.K., T.F.W., L.L. and F.L. analysed the data. J.R., L.M.K.V. and J.S.C. conceived and supervised the project. A.M.J.Z., T.K. and L.M.K.V. wrote the manuscript with input from all authors.\\
\\

\section*{Additional information}
\paragraph*{Data availability} Datasets and analysis scripts supporting the conclusions of this paper are available at https://doi.org/10.5281/zenodo.4478855. 
\paragraph*{Competing interests} The authors declare no competing interests.

\clearpage

\end{document}


\setcounter{page}{13} 

\begin{figure}[h!]
    \centering
    \includegraphics[width=0.5\textwidth]{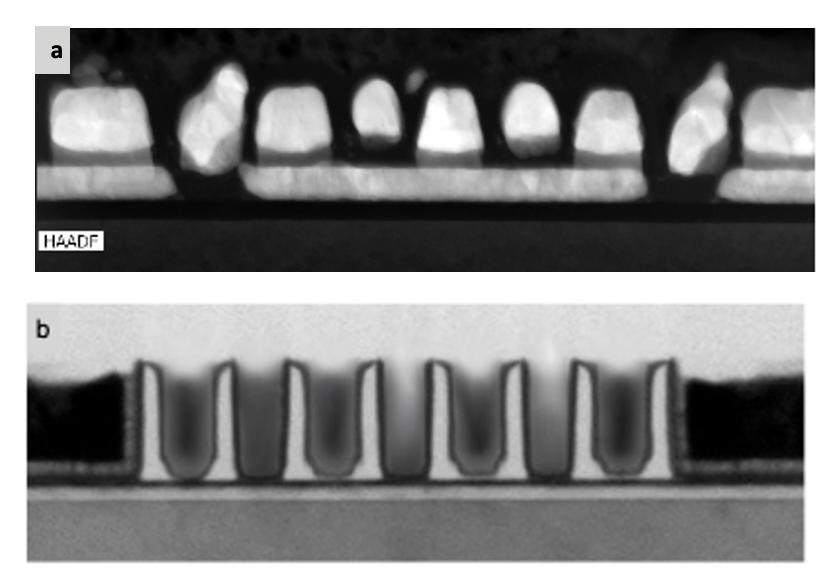}
    \caption{\textbf{Comparison of an academic SiMOS device and a TEM-image of an industrial device.} \textbf{a,} High-angle annular dark-field scanning transmission electron microscopy image (HAADF-STEM) of a SiMOS device from~\cite{eenink_tunable_2019}, fabricated at Delft University of Technology, using electron beam lithography and lift-off techniques. Below the fine gates, the screening gates of the device are visible. \textbf{b,} TEM image of a device nominally identical to devices measured here. Compared to the academic device, the gate profiles and dimensions are much more uniform and well-defined (as stated in the main text, there are two sets of nested gates, so we should compare gates within the same set), and there are no traces of spurious pieces of metal.}
    \label{fig:Ext_data_TEM}
\end{figure}

\clearpage

\begin{figure}[h!]
    \centering
    \includegraphics[width=0.3\textwidth]{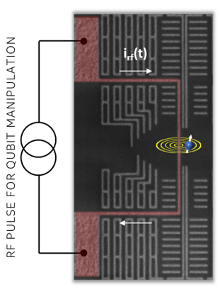}
    \caption{\textbf{Sample with ESR line.} A scanning electron microscope image of a sample with an ESR line, nominally identical to the samples measured in this letter. The ESR line is false-coloured in red. The active area of the sample is indicated by the cartoon of a spin. Metal dummification is clearly visible in the image, the quantum dot gates are not visible as they are covered by dielectric.}
    \label{fig:Ext_data_SEM_stripline}
\end{figure}

\clearpage
\begin{figure}[h!]
    \centering
    \includegraphics{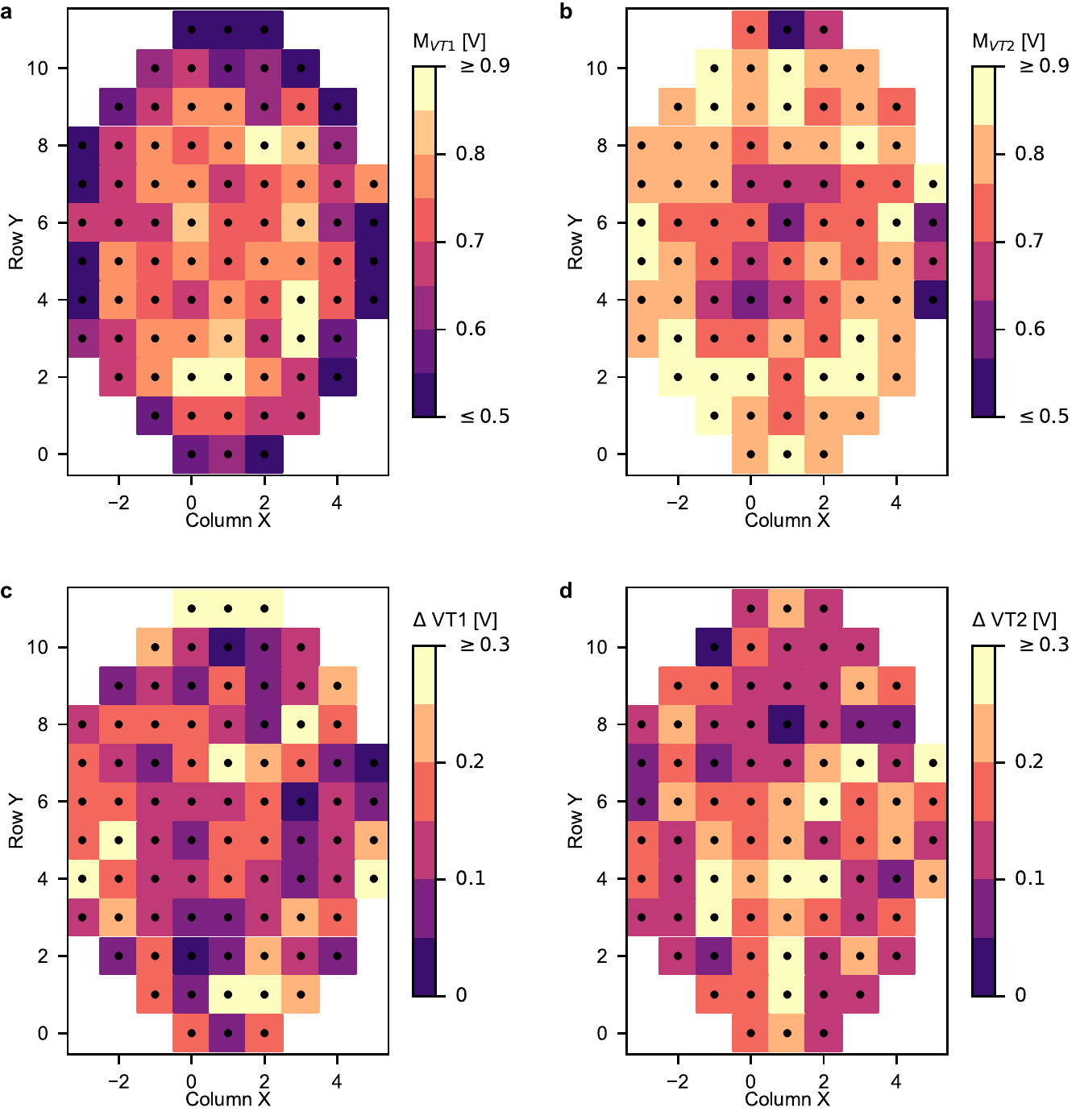}
    \caption{\textbf{Cross-wafer variation of threshold voltages.} Each 300 mm wafer consists of 82 die and each die contains quantum dot arrays with various design skews and array sizes (up to 55 gates), as well as transistor and calibration test structures. To analyse cross-wafer sample uniformity, automated probing at room temperature is used to measure one seven-gate device per die (nominally identical to the devices discussed in the main text). For each device, the threshold voltage for the seven gates on the qubit fin are analysed. \textbf{a,} Median threshold voltage of the three gates of the first gate layer (even-numbered). For the bulk of the samples, the median threshold voltage is between $0.6$~V and $0.8$~V. \textbf{b,} Median threshold voltage for the four gates of the second gate layer (odd-numbered gates). \textbf{c,} Spread (highest minus lowest value) in threshold voltage of the three gates in the first gate layer per sample. For the majority of the wafer, the spread in threshold voltage is between $50$ and $200$~mV. \textbf{d,} Spread in threshold voltage for the second gate layer per sample. In general, the spread in threshold voltage for the majority of the second gate layer is between $100$ and $250$~mV. The combination of figures \textbf{a}, \textbf{b}, \textbf{c} and \textbf{d}. shows that the spread of threshold voltage per sample is low and the median is rather uniform, indicating a good cross-wafer uniformity and thus a precise process control. The cross-wafer variation of the second gate layer is slightly augmented relative to the first layer. This arises from small process-induced cross-wafer variability of the final effective oxide thickness of the second gate layer, which gives rise to an electrostatic change and results in a small shift in threshold voltage.}
    \label{fig:Ext_data_wafer_uniformity}
\end{figure}

\clearpage

\begin{figure}[h!]
    \centering
    \includegraphics[width=0.5\textwidth]{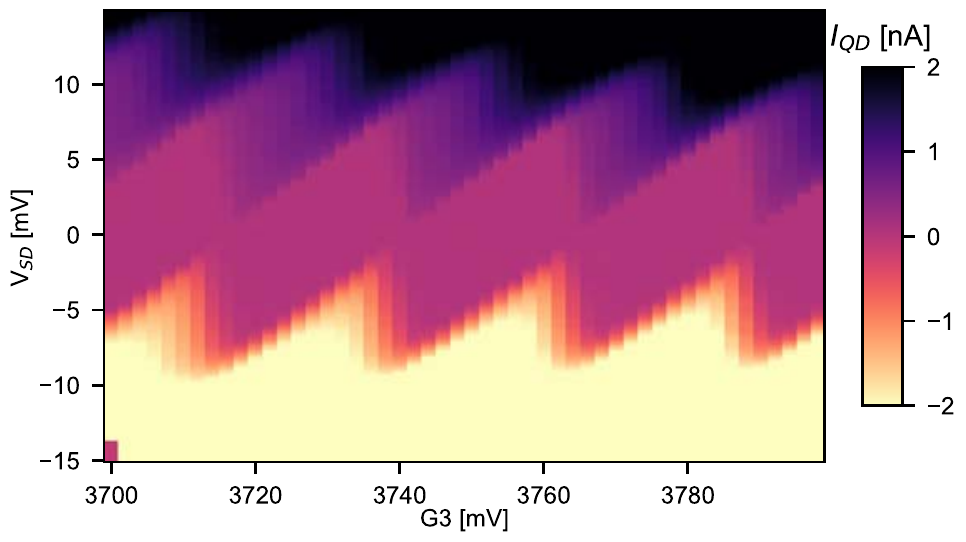}
    \caption{\textbf{Coulomb diamonds} Typical Coulomb blockade diamonds measured in the multi-electron regime, from which the quantum dot charging energy and the gate lever arms are determined. Coulomb diamonds are measured by scanning a gate voltage versus the bias voltage applied between the source and the drain contact. When the electrochemical potential of the quantum dot falls outside the bias window, the current through the quantum dot is blocked, i.e. the sample is in Coulomb blockade. Once the electrochemical potential of the quantum dot is aligned within the bias window, Coulomb blockade is lifted. The bias window increases along the vertical axis, hence Coulomb blockade is lifted over a wider gate voltage range. The regions of blockaded current have the characteristic diamond shape. Here we extract a charging energy of $8.4\pm0.2$~meV and a lever arm of $0.36$~meV/mV.}
    \label{fig:Ext_data_Coulomb_diamonds}
\end{figure}

\clearpage
\begin{figure}[h!]
    \centering
    \includegraphics{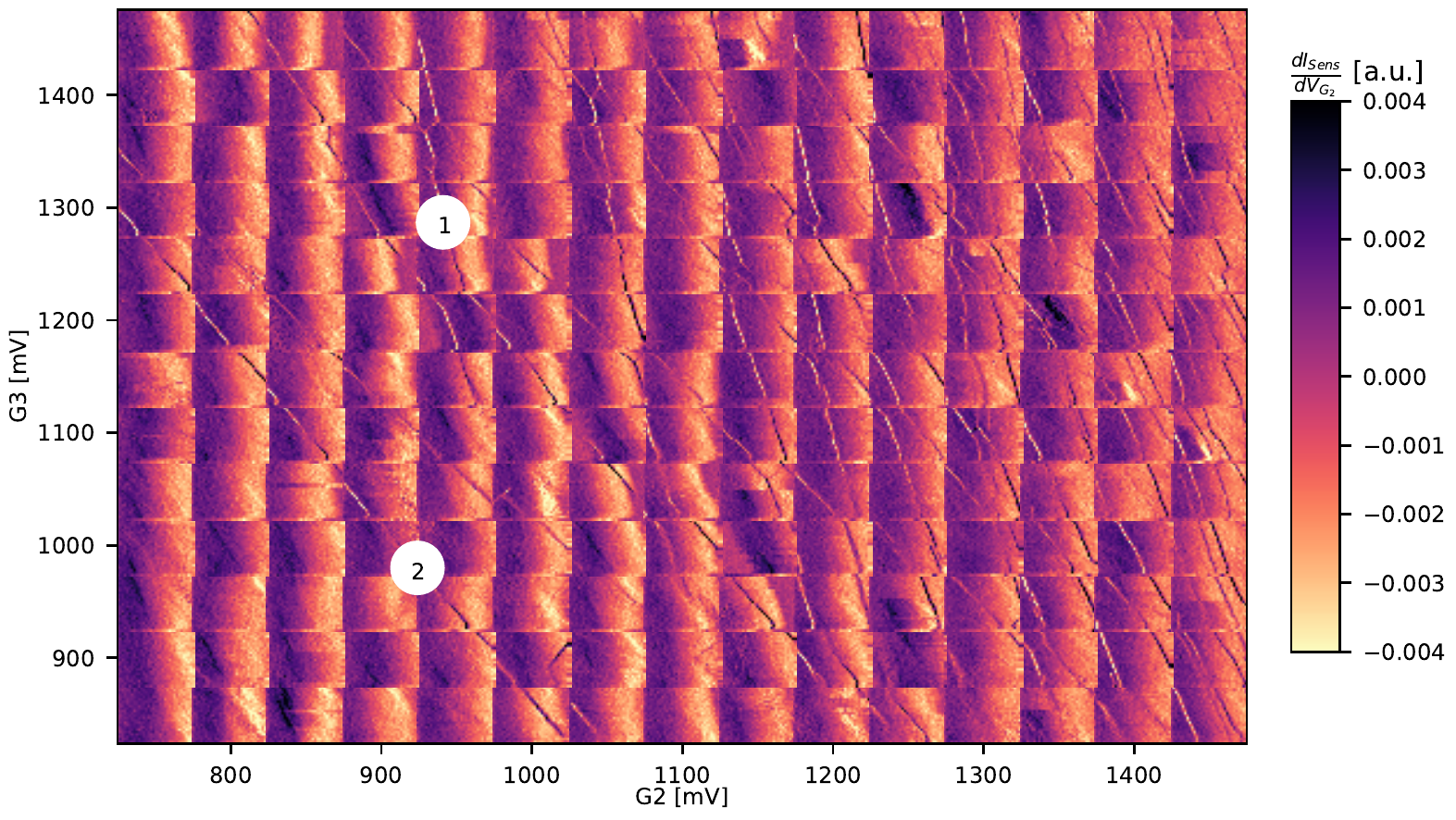}
    \caption{\textbf{Charge stability diagram for the few-electron regime.} Charge stability diagram for the few-electron regime of the samples in which qubit 1 and qubit 2 are measured. Smaller charge sensing maps are stitched together to obtain one large map. As pointed out in the main text, the Coulomb peaks become more irregular towards the single-electron regime, indicating dots forming under adjacent gates. The approximate gate voltages at which qubit 1 and qubit 2 are measured are indicated.}
    \label{fig:Ext_data_large_CSD}
\end{figure}

\clearpage
\begin{figure}[h!]
    \centering
    \includegraphics{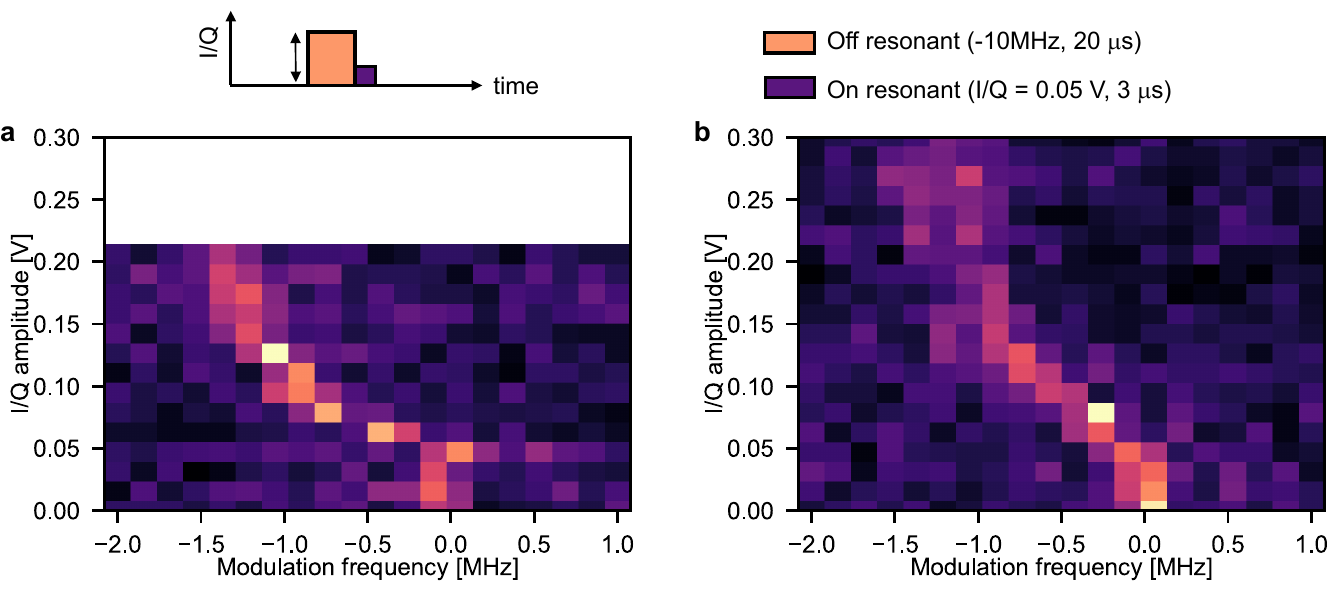}
    \caption{\textbf{Frequency shift due to off-resonant microwave pulse amplitude.} Microwave spectroscopy of \textbf{a,} qubit 1 and \textbf{b,} qubit 2 as a function of the I/Q amplitude of an off-resonant microwave pre-burst (orange in schematic) that is applied immediately before the microwave spectroscopy burst (purple in schematic). Both qubit 1 and qubit 2 show similar behavior with the qubit frequency shifting to a lower frequency when the I/Q amplitude of the off-resonant microwave pre-burst is above $~0.05$~V. The microwave output power at an I/Q amplitude of 0.2~V is $6$~dBm and the LO frequency is $13.072$~GHz for qubit 1 and $13.053$~GHz for qubit 2 (different tuning than in the main text). The off-resonant burst is $10$~MHz away from the LO frequency with a duration of $20$~$\mu$s. The spectroscopic microwave burst has an I/Q amplitude of $0.05$~V with a duration of $3$~$\mu$s.}
    \label{fig:Ext_data_Cryoscope1}
\end{figure}

\begin{figure}[h!]
    \centering
    \includegraphics{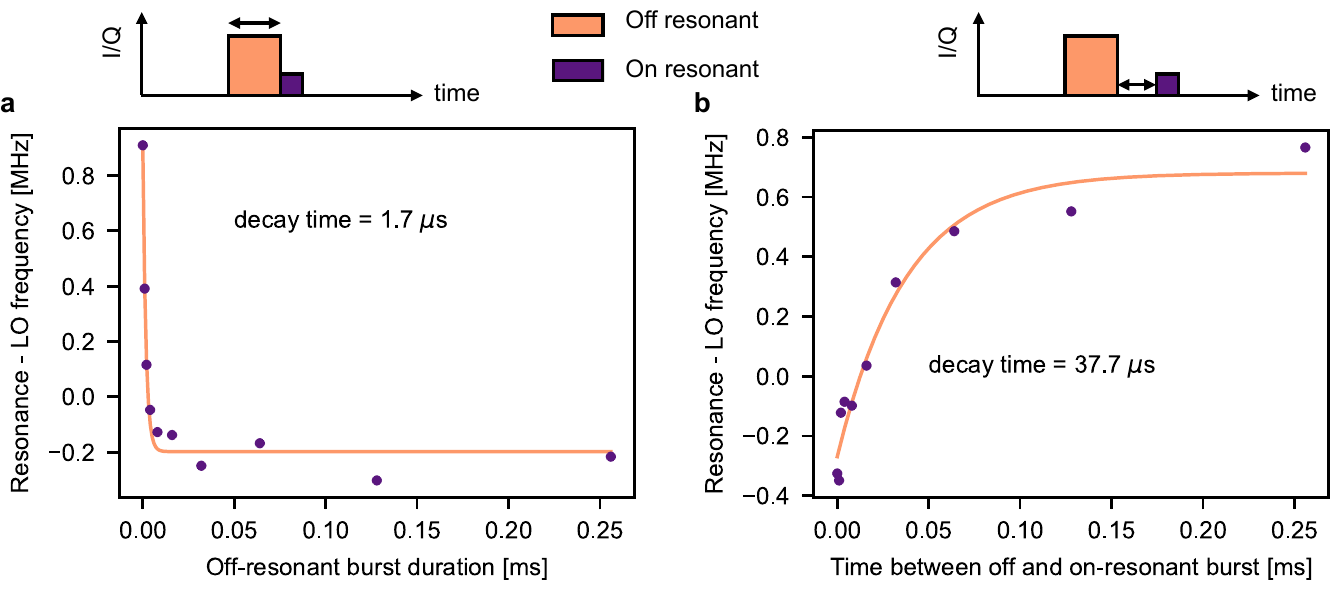}
    \caption{\textbf{Time dependent frequency shift of qubit 1.} In these measurements we perform microwave spectroscopy of the qubit at low power to find the qubit resonance frequency. Before the microwave spectroscopy burst (purple in schematic), we apply an off-resonant burst. \textbf{a,} The resonance frequency of the qubit as a function of the duration of the off-resonant burst applied before spectroscopy (purple dots). An exponential fit gives a time constant of $1.7$~$\mu$s. \textbf{b,} The resonant frequency of the qubit as a function of the time between the off-resonant and spectroscopy pulse. An exponential fit gives a time constant of $37$~$\mu$s. The time dependence of the resonance frequency of the qubit while turning on and off the microwave signal indicates that the frequency shift is related to heating. The off-resonant burst is applied $5$~MHz away from the LO and has an I/Q amplitude of $0.2$~V. The spectroscopy burst has an I/Q amplitude of $0.05$~V and a duration of $3.5$~$\mu$s. The LO frequency is $17.1428$~GHz and the microwave output power is  $13$~dBm at I/Q amplitude of 0.2~V. 
}
    \label{fig:Ext_data_Cryoscope2}
\end{figure}

\clearpage
\begin{figure}[h!]
    \centering
    \includegraphics{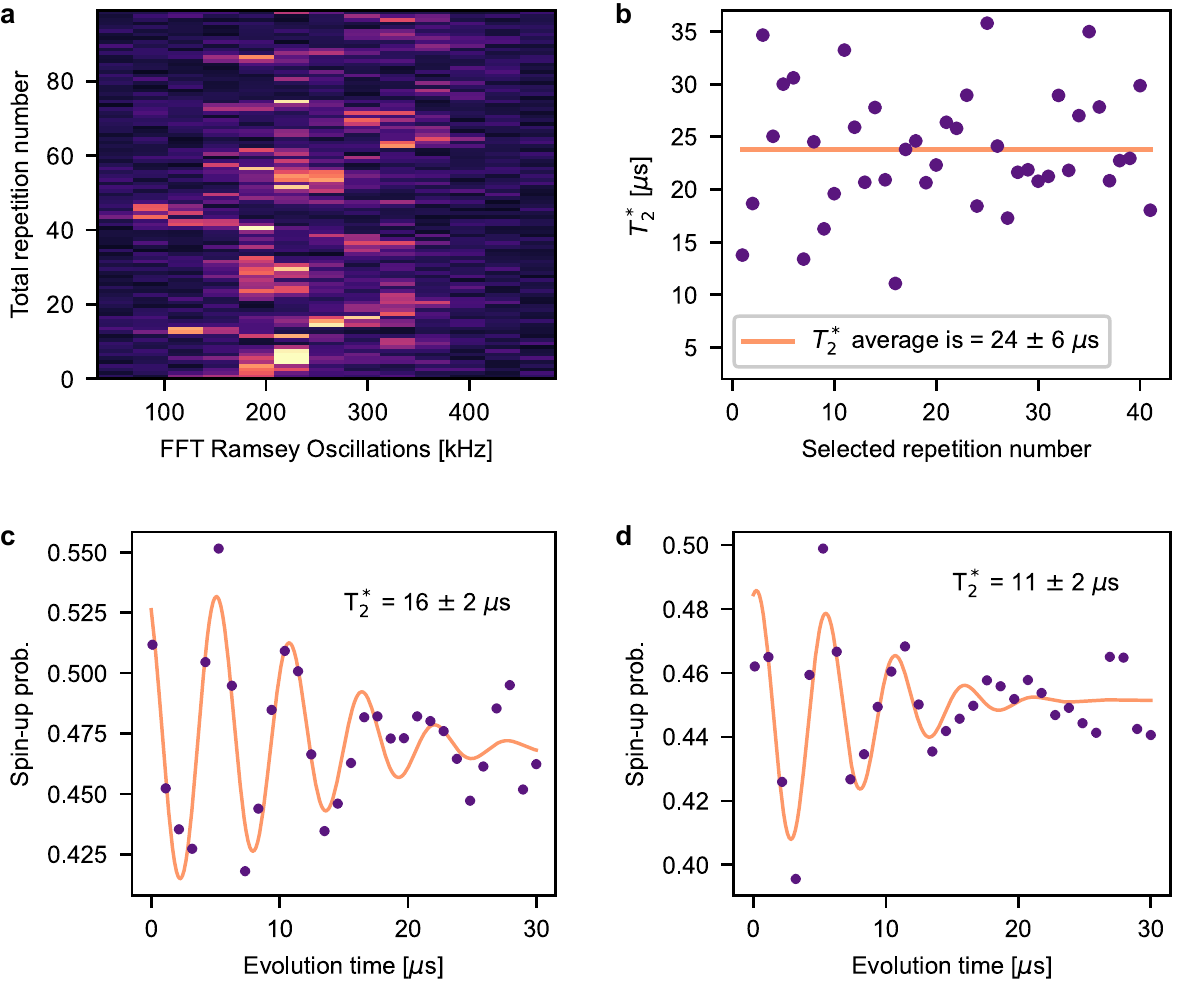}
    \caption{\textbf{Ramsey analysis over time.} \textbf{a,} Fast Fourier transform (FFT) of a Ramsey experiment. The data consists of 100 traces, each trace is an average of 200 repetitions. The entire measurement takes three hours in total. The FFT of the Ramsey fringes shows frequency jumps over the timescale of the measurement. Each trace is fitted individually with a decaying Gaussian curve and the fit is analysed by calculating the mean square error. We keep the 41 traces with mean square error below a given threshold. \textbf{b,} Extracted $T_2^*$ for the selected traces (purple dots). The average of the $T2^*$ times of the selected traces is 24$\pm$6 $\mu$s (orange line). \textbf{c,} Ramsey decay curve. The data points are the averaged data of the 41 selected traces. The fit gives a $T_2^*$ time of 16$\pm$2 $\mu$s. \textbf{d,} Ramsey decay curve. The data points are the average of all 100 traces of the two hour and 40 minute measurement. The fit gives a $T_2^*$ time of 11$\pm$2 $\mu$s.}
    \label{fig:Ext_data_Ramsey}
\end{figure}

\clearpage
\begin{figure}[h!]
    \centering
    \includegraphics[width=0.7\textwidth]{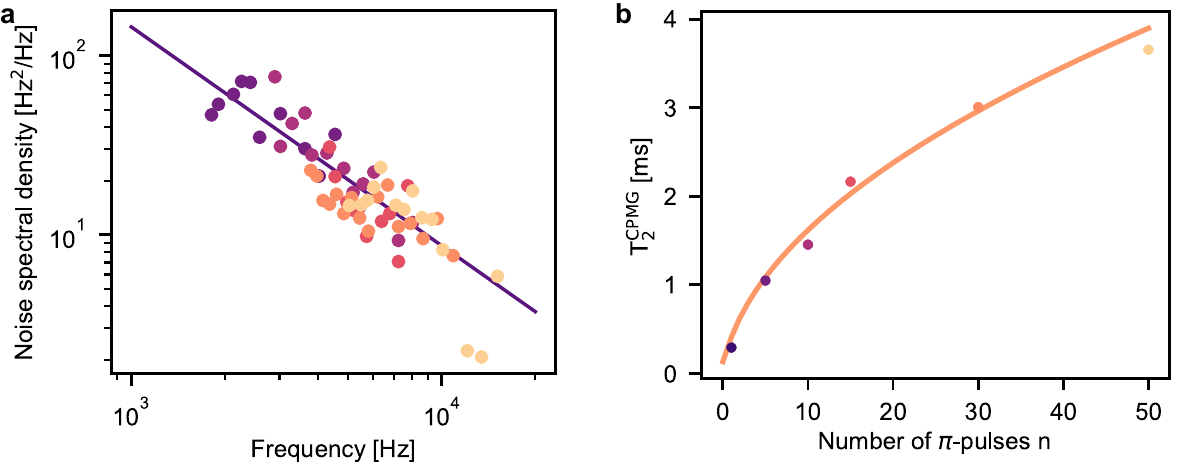}
    \caption{\textbf{Noise analysis for qubit 1.} \textbf{a,} Dynamical decoupling pulses for which the time between two subsequent $\pi$-pulses is fixed, can act as a filter function for Gaussian noise~\cite{bylander_noise_2011, cywinski_how_nodate}. The filter function peaks at a frequency of $f_0=\frac{n}{2t_{wait}}$, with $n$ the number of $\pi$-pulses and $t_{wait}$ the evolution time. When the filter is sufficiently narrow around the frequency $f_0$, the noise within the bandwidth can be regarded as constant. We can use this to relate the amplitude of the CPMG decay, $A_{CPMG}$, for each wait time and number of $\pi$ pulses, to the dominant noise spectrum for the qubit: $S(f_0)=-\ln(A_{CPMG})/2\pi^2t_{wait}$~\cite{kawakami_2016, yoneda_quantum-dot_2018}. Here, we plot the noise spectrum, $S(f)$ as a function of frequency for the data shown in Fig.~4{c} of the main text. We only take data points into account for which $0.15 < A_{CPMG} < 0.85$. We assume that the noise spectrum dominating spin decoherence is described by a power law and use the fit function $Bf^{-\gamma}$. We obtain $\gamma=1.2\pm0.1$ and a noise value at $1$~Hz of $B=820\pm750$~Hz/$\sqrt{\mathrm{Hz}}$. This agrees well with the fitting of the individual CPMG curves and the $T_{2,CPMG}$ scaling in Fig.~4 of the main text. Using the susceptibility of the qubit resonance frequency to a voltage change of a nearby gate and the energy lever arm of that gate we can compare the noise value B to charge noise. With this conversion we obtain $B=29\pm27$~$\mu$eV/$\sqrt{\mathrm{Hz}}$. 
    \textbf{b,} Following ~\cite{bylander_noise_2011}, we now use the filter function as described in \textbf{a} to estimate the noise decay and the noise level at $1$~Hz that gives the given $T_{2, CPMG}$ for the number of $\pi$-pulses and fit this to our data. We obtain $\gamma=1.06\pm0.13$ and $B=30\pm18$~$\mu$eV/$\sqrt{\mathrm{Hz}}$. This is comparable to the results obtained in \textbf{a}.}
    \label{fig:Ext_data_NoisePSD}
\end{figure}

\clearpage

\begin{figure}[h!]
    \centering
    \includegraphics[width=\textwidth]{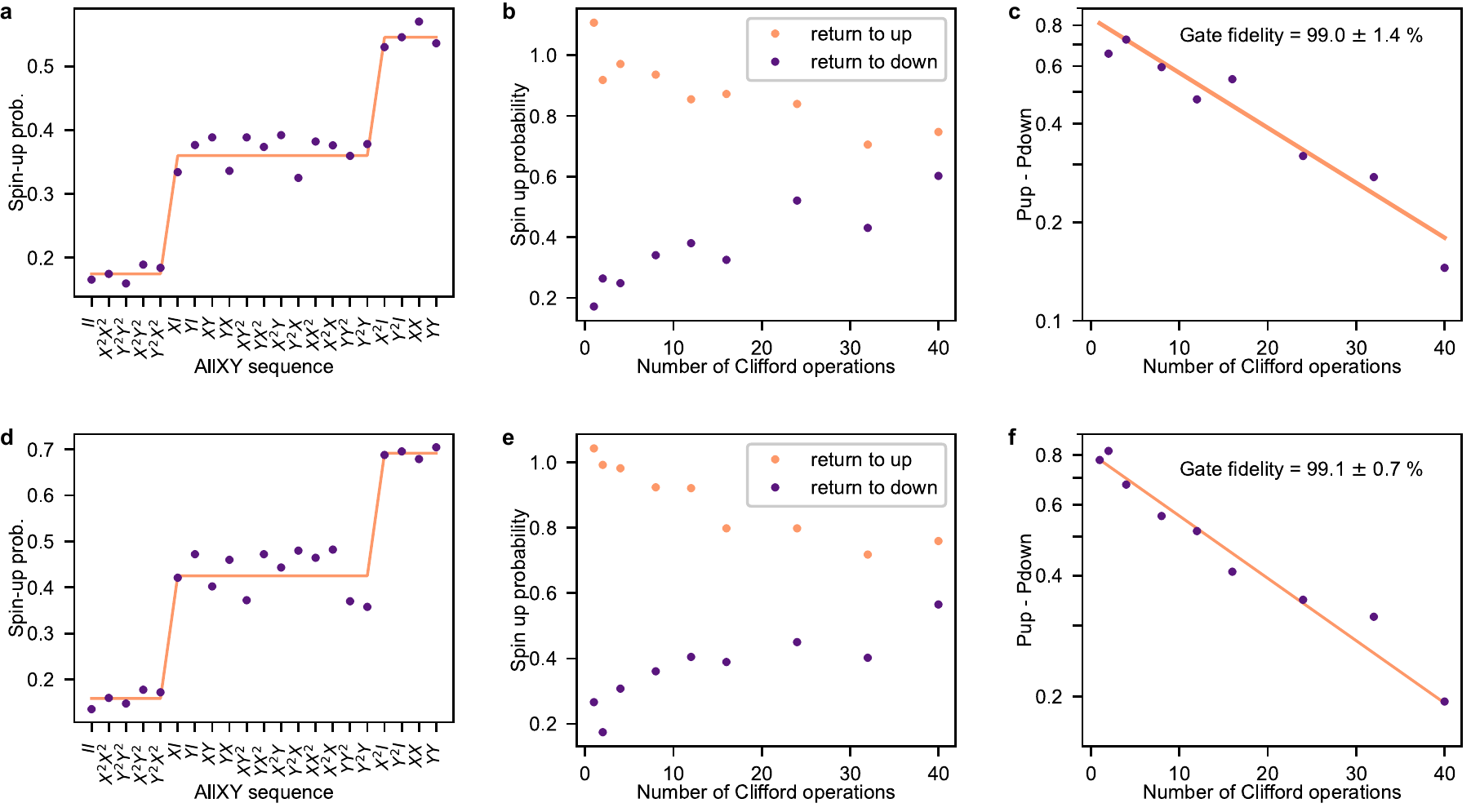}
    \caption{\textbf{AllXY and Randomised benchmarking.}
    \textbf{a, d,} To check the calibration of the single-qubit gates we perform an AllXY sequence~\cite{thesis_reed_2013} on \textbf{a,} qubit 1 and \textbf{d,} qubit 2. Each data point corresponds to the outcome after sequentially applying two gates from the set {I, X, X$^2$, Y, Y$^2$}, where X and Y indicate 90$^{\circ}$ rotations. The data points should ideally follow a staircase pattern (solid line) and deviations from this indicate calibration errors. \textbf{b, c, e, f,} To determine the single-qubit gate fidelity, we perform randomised benchmarking~\cite{knill_randomized_2008, Magesan_characterizing_2012}. In randomised benchmarking, we randomly select $p$ gates from a set of 24 gates that form the Clifford group and apply them to the qubit. At the end of the sequence we apply an inverting gate from the Clifford group that ideally takes the qubit state back to either $\ket{0}$ or $\ket{1}$. In this experiment, the Clifford gates are decomposed to the set of primitive gates {I, $\pm$X, $\pm$X$^2$, $\pm$Y, $\pm$Y$^2$}. On average a Clifford gate contains 1.875 primitive gates. \textbf{b, e,} Normalised spin-up probability as a function of the number of Clifford operations applied for \textbf{b,} qubit 1 and \textbf{e,} qubit 2. The orange and purple data points correspond to sequences producing a net Clifford of X$^2$ or I, respectively, ideally taking the spin to either spin up or spin down. Each data point corresponds to 40 randomisations of the Clifford sequence and the normalisation is done by additional calibration experiments where we apply either just I or X$^2$ to the qubit. \textbf{c, f,} The difference between the purple and orange data points in \textbf{b, e,} is fitted with and exponential of the form $VP_c^p$. From this we derive an average Clifford-gate fidelity of $F_C = 1-(1-P_c)/2$ of $98.1\pm 2.7 \%$ and $98.2\pm 1.4 \%$ for qubit 1 and qubit 2 respectively. This translates to a primitive gate fidelity of $99.0\pm1.4\%$ and $99.1\pm0.7\%$ for Q1 and Q2.}
    \label{fig:Ext_data_RB}
\end{figure}

\clearpage

\begin{figure}[h!]
    \centering
    \includegraphics[width=\textwidth]{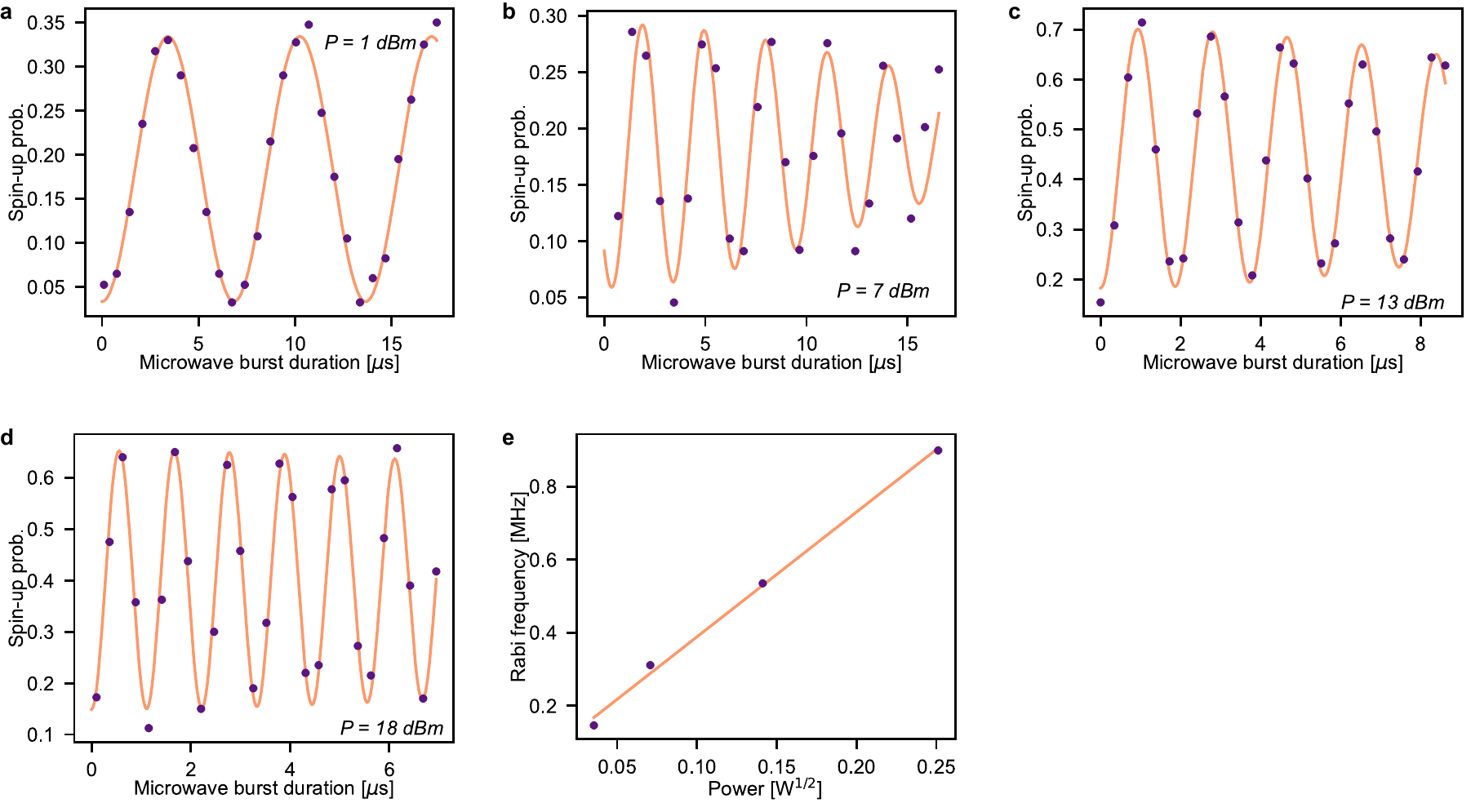}
    \caption{\textbf{Rabi oscillations for qubit 2.} \textbf{a-d,} Rabi oscillation of qubit 2 measured for different output powers of the microwave source. \textbf{e,} The extracted Rabi frequency is plotted versus the square root of the applied power, showing the expected linear dependence.}
    \label{fig:Ext_data_Q2_Rabi}
\end{figure}
\begin{figure}[h!]
    \centering
    \includegraphics[width=\textwidth]{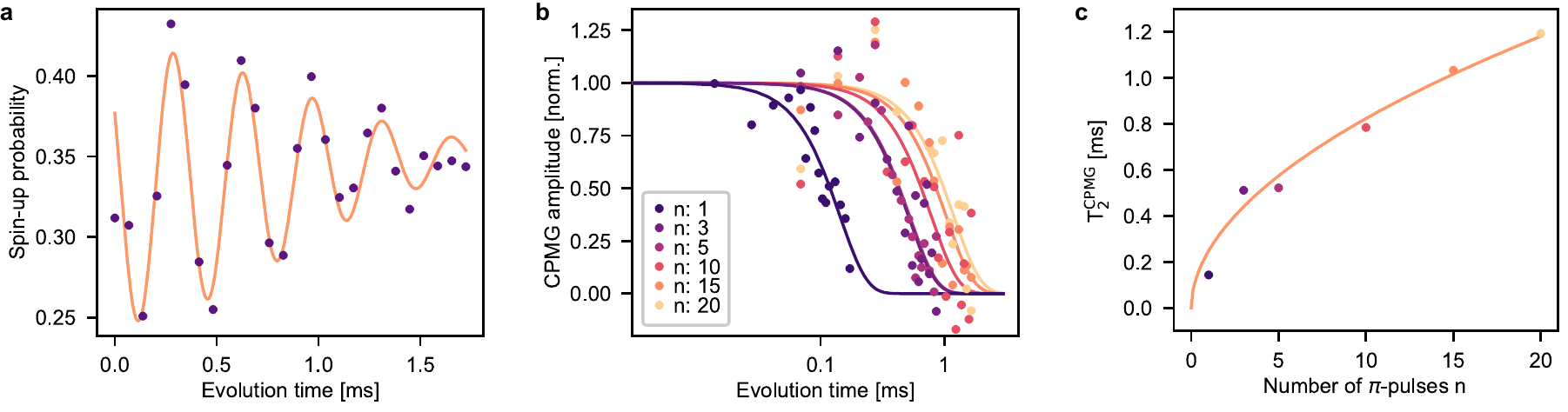}
    \caption{\textbf{Coherence of qubit 2.} \textbf{a,} CPMG-curve for qubit 2 for $n=20$. Fitting this curve, as described in the methods, gives $T_{2,CPMG}=1.2\pm 0.2$ ms. \textbf{b,} Analogously to the case of qubit 1 (see main text), we demodulate and normalise the CPMG amplitude as a function of evolution time for different numbers of $\pi$ pulses, giving the CPMG amplitude. \textbf{c,} The measured CPMG decay time as a function of the number of $\pi$-pulses. The orange line represents a fit through the data (excluding $n=1$) following $T_2^\mathrm{CPMG}\propto n^{(\gamma/(\gamma+1))}$. We extract $\gamma=1.51\pm0.15$. Performing a similar analysis as has been done for qubit 1 (see Extended Data Fig.~\ref{fig:Ext_data_NoisePSD}) gives unreliable results.}
    \label{fig:Ext_data_Q2_CPMG}
\end{figure}

\clearpage
\begin{figure}[h!]
    \centering
    \includegraphics{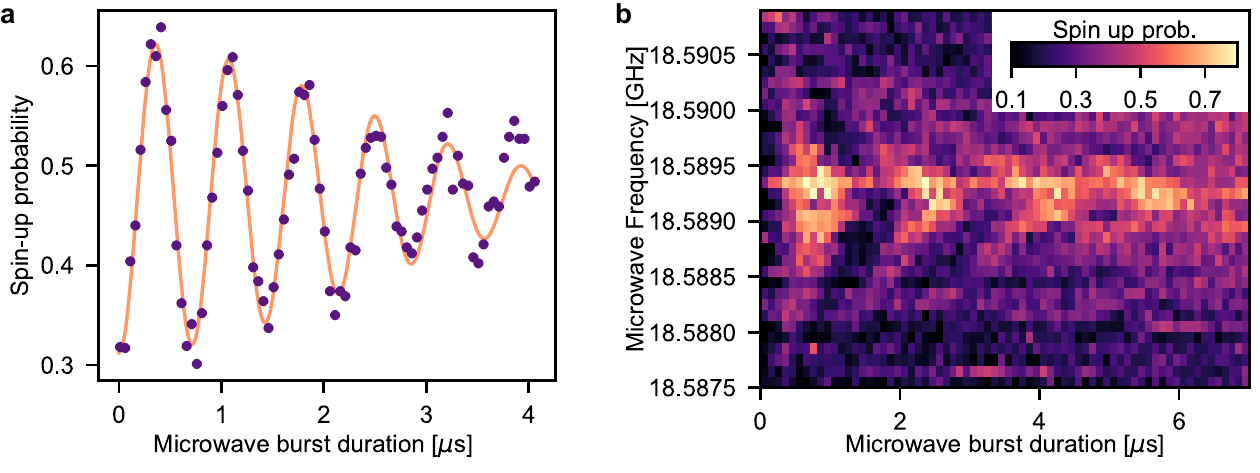}
    \caption{\textbf{Rabi oscillations for qubit 3}. \textbf{a,} Rabi oscillation for a third qubit measured on a different device than qubits 1 and 2. The qubit was measured at an external magnetic field of $B=0.675$~T, giving a Larmor frequency of $18.757$~GHz. From fitting the curve, we extract a Rabi frequency of 1.4~MHz. \textbf{b,} Spin-up probability versus burst duration and microwave frequency in a slightly different tuning regime. The expected Chevron pattern is visible. We observed a second (spurious) quantum dot in the vicinity of qubit 3 and expect that hybridisation with this extra quantum dot is limiting the $T_2^*$ and also the $T_{2,Rabi}$ of qubit 3.}
    \label{fig:Ext_data_Q3_Rabi}
\end{figure}

\bibliography{literature}
\bibliographystyle{naturemag}